\documentclass{sig-alternate-05-2015}

\usepackage{enumitem}
\usepackage{dblfloatfix}
\usepackage{multirow}
\usepackage{footnote}
\usepackage{array}
\usepackage{booktabs}
\usepackage{balance}

\begin{document}

\CopyrightYear{2016}
\setcopyright{acmlicensed}
\conferenceinfo{CSED'16,}{May 14-15 2016, Austin, TX, USA}
\isbn{978-1-4503-4157-8/16/05}
\acmPrice{\$15.00}
\doi{http://dx.doi.org/10.1145/2896941.2896944}
\title{Will My Tests Tell Me If I Break This Code?\titlenote{
This work was partially funded by the German Federal Ministry of Education and Research (BMBF), grant ``Q-Effekt, 01IS15003A''. The responsibility for this article lies with the authors.}}

\numberofauthors{2}

\author{
\alignauthor
Rainer Niedermayr, Elmar Juergens\\
       \affaddr{CQSE GmbH}\\
       \affaddr{Garching b. M\"unchen, Germany}\\
       \email{\{niedermayr, juergens\}@cqse.eu}
\alignauthor 
Stefan Wagner\\
       \affaddr{University of Stuttgart}\\
       \affaddr{Stuttgart, Germany}\\
       \email{stefan.wagner@informatik.uni-stuttgart.de}
}

\maketitle

\begin{abstract}
Automated tests play an important role in software evolution
 because they can rapidly detect faults introduced during changes.
In practice, code-coverage metrics are often used as criteria to evaluate the effectiveness of test suites
 with focus on regression faults.
However, code coverage only expresses which portion of a system has been executed by tests,
 but not how effective the tests actually are in detecting regression faults.

Our goal was to evaluate the validity of code coverage as a measure for test effectiveness.
To do so, we conducted an empirical study in which we applied an extreme mutation testing approach
 to analyze the tests of open-source projects written in Java.
We assessed the ratio of pseudo-tested methods
 (those tested in a way such that faults would not be detected)
 to all covered methods and judged their impact on the software project.
The results show that the ratio of pseudo-tested methods is acceptable for \textit{unit} tests
 but not for \textit{system} tests (that execute large portions of the whole system).
Therefore, we conclude that the coverage metric is only a valid effectiveness indicator for unit tests.

\end{abstract}

% http://dl.acm.org/ccs.cfm




This code snippet demonstrates that high code coverage does not imply test effectiveness:

\begin{verbatim}
public class Calculation {
  private int value;
		
  public Calculation() {
    this.value = 0;
  }
  public void add(int x) {
    this.value += x;
  }
  public boolean isEven() {
    return this.value % 2 == 0;
  }
}
\end{verbatim}

The class \texttt{Calculation} consists of an integer field named \texttt{value} and two public methods.
The \texttt{add} method allows for adding an integer to the internal value.
The \texttt{isEven} method returns a boolean value, which indicates if the current internal value is an odd or even number.
This class is tested by a JUnit test in the \texttt{CalculationTest} class.

\begin{verbatim}
public class CalculationTest {
  @Test
  public void testCalculation() {
    Calculation calc = new Calculation();
    calc.add(6);
    assertTrue(calc.isEven());
  }
}
\end{verbatim}

The test case creates a new instance of the \texttt{Calculation} class and implicitly assigns 0 to the field \texttt{value}.
It then uses the \texttt{add} method to increase the internal value by 6.
Finally, the test case verifies that the \texttt{isEven} method returns true, which is expected for 6.

The test case executes all methods, statements, and bran\-ches of the class under test.
This results in 100\% code coverage at the method, statement and branch levels.
Consequently, one could assume that the \texttt{Calculation} class is effectively tested.

However, this is not the case.
Let's assume that the programmer forgot to implement the logic of the \texttt{add} method so that its body is empty. 
Although the test case covers that method, it will not detect the fault,
 because both 0 and 6 are even numbers.
Therefore, the \texttt{add} method is executed, but not effectively tested, because the test case would not detect any faults.
We call this a pseudo-tested method.

According to Fowler \cite{wwwFowler}, test cases that do not contain any assertions are another example of tests
 that increase the coverage, but are useless (unless their purpose is to check if exceptions are thrown).

\section{Fundamentals and Terms}
\label{Section_Fundamentals}
This section describes the terms used in this paper.

\vspace{\baselineskip}
A \textit{unit test} examines a small unit of code (usually a method or a class).
It consists of a sequence of method calls and assertions that verify that the computed results of the invocations equal the expected ones.
A unit test can also check the absence of thrown exceptions for a given program flow.

\vspace{\baselineskip}
A \textit{system test} examines a complete software system, which may consist of several components.
Unlike a unit test, it covers many methods by executing a large proportion of the whole system.
A system test often triggers the execution of a large functionality
 and compares the end result (which can be aggregated data, a report, or a log file) with the expected one.

\vspace{\baselineskip}
\textit{Code coverage} is a metric that expresses which ratio of application code of a software project is executed when running all test cases.
It can be computed at different levels; widely used are measures at method/function, statement or branch level.
When we refer to code coverage in this paper, we mean method coverage.
It corresponds to the ratio of methods that are executed by tests.

\vspace{\baselineskip}
We consider a method to be \textit{test-executed} if it is covered by at least one test case.
A test-executed method is considered \textit{tested}
 if at least one covering test case fails when the whole logic of the method is removed.
In contrast, a test-executed method is considered \textit{pseudo-tested}
 if none of the covering test cases fails when the whole logic of the method is removed.

\vspace{\baselineskip}
We define \textit{test effectiveness} as the capability of test cases to detect regression faults in methods that they execute.
Intuitively, we understand the overall ratio of code that is effectively tested as the upper bound
 of the probability that test cases detect a novel regression fault in a project.
\section{Related Work}
This paper is based on the master's thesis of the first author \cite{niedermayr2013thesis}.
Related work is in the areas of code coverage, test suite effectiveness and mutation testing.

\subsection{Code Coverage, Test Suite Effectiveness}
Wong, Horgan, London and Mathur \cite{wong1994effect} showed that the correlation between fault detection effectiveness and block coverage
 is higher than between effectiveness and the size of the test set.
This indicates that coverage can be a valid measure for test effectiveness.
However, they did not differentiate between different test types.

In \cite{andrews2006using}, Andrew, Briand, Labiche and Namin conducted an empirical study on one industrial program with known faults
 to investigate test coverage criteria on fault-detection effectiveness.
Their results showed that no one coverage criteria is more cost-effective than any other,
 but more demanding criteria lead to larger test suites that detect more faults.
In contrast, we analyzed open-source systems and made use of mutation testing.

In \cite{mockus2009test}, Mockus, Nagappan and Dinh-Trong revealed that an increase in coverage exponentially increases the test effort
 and linearly reduces the field problems.
They suggested that ``code coverage is a sensible and practical measure of test effectiveness'', but did not differentiate between unit and system tests.

Namin and Andrews \cite{namin2009influence} studied the relationship between size, coverage, and fault-finding effectiveness of test suites.
They found that both size and coverage are important for effectiveness and suggested a nonlinear relationship between them.
They analyzed very small C programs (the largest one consisted of less than 6,000 lines of code).

Inozemtseva and Holmes \cite{inozemtseva2014coverage} came to different conclusions.
They evaluated the relationship between test suite size, coverage, and effectiveness for Java programs.
They found that: ``High levels of coverage do not indicate that a test suite is effective.''
In addition, they discovered that the type of coverage had little effect on the correlation between coverage and effectiveness.
They also used mutation testing, but they generated test suites of a fixed size by randomly selecting test cases.

Marick \cite{marick1999misuse} critically analyzed code coverage as a metric for test effectiveness.
He showed how code coverage is commonly misused and argued for a cautious approach to the use of coverage.
His work did not assess the validity of code coverage as effectiveness indicator.

\subsection{Mutation Testing}
\label{Subsection_Related_Work_Mutation_Testing}
Mutation testing was first proposed in the 1970s and is an established, powerful technique to evaluate test suites.
The general principle is to generate mutants by introducing faults into a program
 and check if the tests can detect (\textit{kill}) these.
Experimental studies
 (\cite{andrews2005mutation}, \cite{daran1996software}, \cite{hutchins1994experiments}, \cite{offutt1996experimental})
 provide evidence that mutation testing is a good indicator for the fault-detection capability of test suites.

Mutation testing has two major drawbacks, which explain why it is not widely used in practice.
First, the computational costs are high because mutation testing involves creating a large number of mutants
 and the execution of all tests that cover the mutated code chunk for each mutant.
Second, some of the generated mutants are semantically equivalent to the original code.
The so-called \textit{equivalent mutants} do not represent injected faults,
 cannot be killed by tests and so distort the results.
The detection of these mutants is generally undecidable.
Gr\"un, Schuler and Zeller \cite{grun2009impact} identified equivalent mutants as an important problem
 because they are surprisingly common.
Our approach addressed these issues and is explained in the next section.
\section{Mutation Approach}
\label{Section_Mutation_Approach}
The general idea is to apply mutation testing as mean to collect data for assessing test effectiveness.
We used an extreme mutation operator in which we eliminated the whole logic from a method
 and determined whether or not a method is pseudo-tested.
Therefore, we got around the two drawbacks of mutation testing (mentioned in Section \ref{Subsection_Related_Work_Mutation_Testing}).
First, we created, at most, two different mutants for a method, keeping the number of mutants manageable
 and the execution time for analyzing a medium-sized software project within a few hours.
Second, the mutation operator radically changes the methods and generates less equivalent mutants.
The majority of the generated equivalent mutants can be identified automatically.

\begin{figure}
	\centering
	\includegraphics[width=0.9\linewidth]{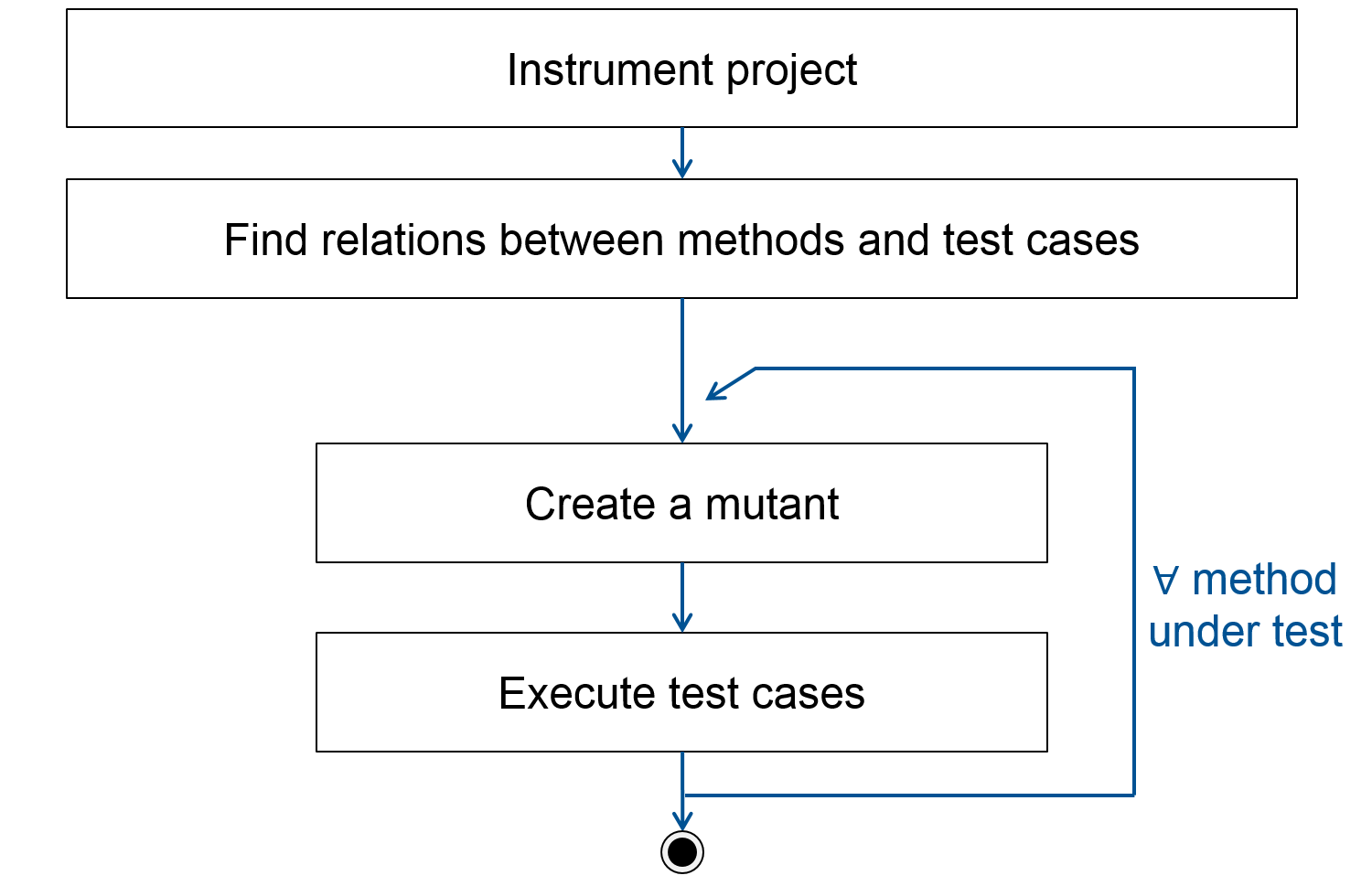}
	\caption{Steps of the mutation analysis process.
	 The first two steps are executed once, the latter ones are executed for each considered method.
	}
	\label{Figure_Mutation_Process}
\end{figure}

\vspace{\baselineskip}
Our approach consists of four steps and is depicted in Figure \ref{Figure_Mutation_Process}.
The first two steps are executed once and are necessary to collect information about the test cases.
The latter two steps comprise the actual mutation process.
They are executed for each method under test and can be run concurrently.
\\
In \textit{Step 1}, the project code is instrumented.
This is done by inserting logging statements.
\\
In \textit{Step 2}, all test cases are executed once on the instrumented code.
This allows us to determine the relationships between test cases and methods.
From the information of the test-executed methods for a given test case,
 the opposite direction of the relation (all test cases covering a given method) can be computed. 
Test cases that fail at this point are excluded from further analysis.
\\
In \textit{Step 3}, the mutation of one method under test is performed.
The mutation operator removes the whole logic of the method.
This is done as follows:

\begin{itemize}[noitemsep]
	\item For void methods, all statements are removed.
	No further actions are necessary.
	\item For methods that return a primitive or a string value, two mutants are generated.
	The mutation removes the original code by replacing it with a return statement.
	The value to be returned depends on the return type and is different for both created mutants.
	Table \ref{tbl:Simple_Return_Value_Generators} lists the return values for the primitive types and string.
	Note that it is sufficient for a method to be considered as tested
	 if at least one of the two mutants can be killed by any of the test cases.
	\item For methods that return an object, a factory is used to generate a suitable instance.
	The original code is removed and the generated instance is returned.
	We developed factories for three study objects (see Section \ref{Subsection_Experiment_Procedure}).
\end{itemize}

\begin{table}
\small
\centering
\caption{Return values for primitive types and string}
\begin{tabular}{lll}
\toprule
Return Type 	          & Mutant 1          & Mutant 2 \\
\midrule
 boolean 								& \texttt{false} 		& \texttt{true} \\
 byte, short, int, long & \texttt{0} 				& \texttt{1} \\
 float, double 					& \texttt{0.0} 			& \texttt{1.0} \\
 char 									& \texttt{' '} 			& \texttt{'A'} \\
 string 								& \texttt{""} 	  	& \texttt{"A"} \\
\bottomrule
\end{tabular}
\label{tbl:Simple_Return_Value_Generators}
\end{table}

A method is excluded from the mutation analysis if it:
\begin{itemize}[noitemsep]
	\item does not contain any statements (otherwise an equivalent mutant would be generated).
	\item is a setter or getter method consisting of exactly one statement (we consider such a method as too trivial to test).
	\item is a constructor.
	\item is a special method generated by the compiler (such as Java synthetic methods).
	\item returns an object but no factory is provided, or the factory cannot create an appropriate instance.
\end{itemize}

In \textit{Step 4}, all test cases that cover the mutated method are executed against the mutated code.
The outcome of this step shows which test cases fail and which still succeed after the mutation.

\section{Experimental Design}
We analyze open-source projects to investigate pseudo-testing
 and the validity of the code-coverage metric at the method level.

\subsection{Study Objects}
To perform the experiment, we chose 14 open-source pro\-jects.
The projects satisfied the following criteria: 
They must use Java as a programming language because the mutation approach is performed on the Java bytecode.
Moreover, the projects must contain tests that use either the JUnit or testNG framework.

The selected projects are of different sizes.
The smallest one consists of about 7,000 lines of code (LOC),
 and the largest project measures about 500,000 LOC.
The projects can be classified into libraries with unit tests
 and systems with system tests.
For systems, we considered only system tests, even if they also had unit tests.
Table \ref{Overview_Study_Objects} lists the projects and their characteristics.

\begin{savenotes}
\begin{table}
  \small
	\centering
	\caption{Study Objects}
	\begin{tabular}{lrrl}
	\toprule
	 Name & LOC & Tests & Test Type \\
	\midrule
	 Apache Commons Collections & 109,415 & {4,468} & {unit} \\ 
	 Apache Commons Lang &  100,422 & {2,000} & {unit} \\
	 Apache Commons Math & 275,570 & {3,463} & {unit} \\
	 Apache Commons Net & 53,601 & {133} & {unit} \\
	 ConQAT Engine Core & 27,481 & {107} & {unit} \\
	 ConQAT Lib Commons & 43,419 & {468} & {unit} \\
	\midrule
	 ConQAT dotnet & 8,239 & {22} & {system} \\
	 DaisyDiff & 11,288 & {1\footnote{The single system test is parameterized and executed with 247 different input files.}} & {system} \\
	 Histone & 24,412 & {93} & {system} \\
	 LittleProxy & 7,300 & {93} & {system} \\
	 Predictor & 7,733 & {21} & {system} \\
	 Struts 2 & 148,486 & {6} & {system} \\
	 Symja & 443,092 & {448} & {system} \\
 	 Tspmccabe & 44,999 & {10} & {system} \\
	\bottomrule
	\end{tabular}
	\label{Overview_Study_Objects}
\end{table}
\end{savenotes}

\subsection{Research Questions}
In this experiment, we wanted to determine the validity of code coverage as a criterion for test effectiveness.
The effectiveness expresses how likely it is that newly introduced faults in test-executed methods are revealed by test cases.

\vspace{\baselineskip}
\noindent
\textbf{RQ 1: What is the ratio of pseudo-tested methods?}
\\
\indent
Research Question 1 evaluates how many methods are test-executed but are actually only pseudo-tested.
As pseudo-tested methods contribute to the overall code coverage of a software project,
 the amount of test-executed code suggested by the code-coverage metric is higher than the amount of (effectively) tested code.
Therefore, code coverage might not be a valid indicator for test effectiveness.

\vspace{\baselineskip}
\noindent
\textbf{RQ 2: Does the ratio of pseudo-tested methods depend on the type of test?}
\\
\indent
Research Question 2 investigates the influence of the type of test (unit tests versus system tests) 
 on the ratio of pseudo-tested methods.
We expect methods that are test-executed by system tests to more likely be pseudo-tested
 because system tests execute many methods in one run.
Therefore, we want to find out if code-coverage validity should be assessed separately for these two types of tests.

\vspace{\baselineskip}
\noindent
\textbf{RQ 3: How severe are the pseudo-tested methods?}
\\
\indent
Research Question 3 analyzes the functional purpose and the severity of pseudo-tested methods.
We want to understand how severe is the lack of test effectiveness of these methods for a project.
Some methods may not warrant being tested (because they are too trivial or non-deterministic),
 but others may be relevant and need more thorough testing to detect regression faults.

\subsection{Experiment Design}
First, we ran the mutation testing analysis for the study objects using our Java program.
It already comprised the exclusion of provable equivalent mutants and simple setters and getters
 (see Section \ref{Section_Mutation_Approach}).
Then, we analyzed the obtained data to answer the research questions.
The design for each research question is as follows:

\vspace{\baselineskip}
\noindent
\textbf{RQ 1.}
We denoted the number of pseudo-tested methods as \(M_{PT}\)
 and calculated the ratio of pseudo-tested methods as follows:
\begin{center}
$ r(M_{PT}) = \frac{\text{number of pseudo-tested methods (} M_{PT} \text{)}}{\text{number of mutated test-executed methods}} $
\end{center}
The mutated test-executed methods comprise all methods
 for which a mutant is created and tested (by at least one test case without timeout) during the mutation analysis
 (refer to Section \ref{Section_Mutation_Approach} for excluded methods).
\\
Consequently, the ratio of tested methods is:
\begin{center}
$ r(M_{T}) = 1 - r(M_{PT}) $
\end{center}
We also listed the method coverage (\(CC\)) of the study objects
 and computed the overall ratio of tested methods to all existing methods:
\begin{center}
$ r(CT) = CC * r(M_{T}) $
\end{center}
(assuming that the test-executed methods that were excluded from the mutation analysis provide an approximately similar result).
We considered \(r(CT)\) as an upper bound for the ratio of tested code to the whole project.
Test cases can find faults in this portion of the project.

\vspace{\baselineskip}
\noindent
\textbf{RQ 2.}
To answer this question, we examined the ratio of pseudo-tested methods for unit and system tests separately.
We then compared the mean ratio of pseudo-tested methods per test type, computed the standard deviation, and compared the data using boxplots.

\vspace{\baselineskip}
\noindent
\textbf{RQ 3.}
We investigated all pseudo-tested methods and assigned each one to a functional category.
The mapping was done by manual inspection based on the method name.
We then assigned a severity to each functional category.
Table \ref{Method_Categories} presents the categories and their severities.
The severities are defined as follows:

\begin{itemize}[noitemsep]
		\item Functional categories with methods that are not deterministic (such as generating a random number) or not intended to be tested
		       were assigned a severity of \textit{irrelevant}.
					\\
					Moreover, unless the hash generation is explicitly test\-ed, the \texttt{hashcode()} method belonged to the severity irrelevant
					 because the logic of this method still corresponds with its specification after the mutation if it always returns the same constant value.
		\item The severity \textit{low} is assigned to functional categories that contain methods
		       that do not significantly influence the program execution.
					The most prominent examples are validation and optimization methods, as well as those to close streams and connections.
		\item The severity \textit{medium} was for categories with methods that are likely to influence the program execution to some extent.
					These include methods that send events to listeners, prepare a computation, or set or get properties\footnote{
					Note that our analysis results do not contain very simple setters and getters consisting of a single statement
					(as described in Section \ref{Section_Mutation_Approach}).},
					 or transform or convert a value.
					We also assigned the \texttt{toString()} method that returns a string representation of an object to a severity of \textit{medium}.
		\item The severity \textit{high} was assigned to categories that contain methods likely to be very relevant for the program execution.
					They concern the computation logic or important data structures.
					Examples are: \texttt{calculateAST()}, \texttt{writeToFile()}, \texttt{storeObject()}.
					\\
					Moreover, this severity comprises the \texttt{equals(Object)} method, which checks if an object is semantically equal to another one,
					 and the \texttt{compareTo(T)} method, which returns the natural order of two objects.
	\end{itemize}
	
\begin{table}
  \small
	\centering
	\caption{Functional categories and their severity}
	\begin{tabular}{llp{3.2cm}}
\toprule
	 Functional Category & Severity & Examples \\
\midrule
 	 hashcode            & irrelevant & \texttt{hashcode()} \\
	 non-deterministic   & irrelevant & \texttt{setSeed(int)} \\
	 test-related        & irrelevant & \texttt{updateTestData()} \\
\midrule
	 finalization        & low        & \texttt{finalize()}, \mbox{\texttt{closeStream()}} \\
	 monitoring          & low        & \texttt{logInfo(String)} \\
	 optimization        & low        & \texttt{addToCache(Object)} \\
	 validation          & low        & \texttt{checkIndex(int)}, \mbox{\texttt{validateParam(Object)}} \\
\midrule
	 events              & medium     & \texttt{notifyListeners()}, \mbox{\texttt{firePropertyChange()}} \\
	 preparation         & medium     & \texttt{initWorkflow()}, \mbox{\texttt{setUpBlock()}} \\
	 setter and getter   & medium     & \texttt{isRed(Color)}, \mbox{\texttt{getV(int)}} \\
	 toString            & medium     & \texttt{toString()} \\
	 transformation      & medium     & \texttt{abs(int)}, \mbox{\texttt{escape(String)}} \\
\midrule
	 object identity     & high       & \texttt{equals(Object)}, \mbox{\texttt{compareTo(T)}} \\
	 program logic       & high       & \texttt{computeLSB()}, \mbox{\texttt{solvePhase1()}} \\
\bottomrule
	\end{tabular}
	\label{Method_Categories}
\end{table}

\subsection{Experiment Procedure}
\label{Subsection_Experiment_Procedure}
This section describes the mutation testing analysis execution for the study objects.

We checked out the source code of the projects from the repositories;
 built the code with Ant, Maven, or Gradle according to the instructions;
 and imported the projects in the Eclipse IDE.
We then looked at the source code to select the unit or system tests.
We exported the compiled application and test code as separate jar files and provided the necessary test data.
We specified the location of the jar files and other dependencies in a configuration file.
Moreover, we defined timeouts for the test cases (twice as long as the longest duration of any test case on the original code)
 and the number of concurrently running tests.
Some study objects required the development of a tailored test runner to locate the test cases and run them piecewise.

We then executed the mutation testing analysis for methods with void, primitive or string as return types.
We analyzed the method signatures of the study objects and discovered that about 50\% to 65\% of the methods belonged to this group.
We also developed factories to create instances for three study objects (\textit{Apache Commons Lang}, \textit{Apache Commons Collections}, and \textit{ConQAT Engine Core})
 to investigate methods that return objects in a separate analysis.
Since the gained results were approximately comparable to the results of methods with a primitive return type,
 we did not further investigate methods returning objects in this experiment.

After the completion of the mutation analysis, we looked at the log file to check for any unexpected problems.
If serious problems were logged, we fixed the cause and restarted the analysis.
Finally, we imported the analysis results in the form of SQL statements into a database and stored additional execution information (such as the duration).

Our tool developed to execute the mutation analysis is available on GitHub\footnote{\url{https://github.com/cqse/test-analyzer}}.
\section{Results}
The answers to the research questions suggest that code coverage is a valid effectiveness indicator for unit tests.
This does not apply to system tests because the ratio of pseudo-tested methods is higher for this type of test and heavily deviates, depending on the project.

\vspace{\baselineskip}
\noindent
\textbf{RQ 1: What is the ratio of pseudo-tested methods?}
\\
\indent
The ratio of pseudo-tested methods varied heavily and ranged between 6\% and 53\% for most study objects.
According to the results, the \textit{Apache Commons Lang} methods are mostly effectively tested
 (according to the definition in Section \ref{Section_Fundamentals})
 because the code coverage at the method level was 93\% and less than 2\% of the test-executed methods were classified as pseudo-tested.
Other study objects exhibited a much higher ratio of pseudo-tested methods,
 including the \textit{Predictor}, \textit{Struts 2}, and \textit{LittleProxy} projects with more than half of the test-executed methods being pseudo-tested.

Table \ref{Table_Results_Pseudo_Tested} presents the number \(M_{PT}\) and ratio \(r(M_{PT})\) of pseudo-tested methods,
 the method coverage \(CC\), and the overall ratio of tested methods \(r(CT)\) for each study object.

\begin{table}
\small
\centering
\caption{Overview of results}
\begin{tabular}{llrrrr}
\toprule
Study Object 	  & \(M_{PT}\) & \(r(M_{PT})\) & \(CC\) & \(r(CT)\) \\
\midrule
	Apache Comm. Coll.         & 124 &  9.5\%  & 81.6\% & 73.9\% \\
	Apache Comm. Lang			  	 &  22 &  1.9\%  & 93.0\% & 91.3\% \\
	Apache Comm. Math		  		 & 271 & 10.6\%  & 84.8\% & 75.8\% \\
	Apache Comm. Net		  		 &  28 & 18.4\%  & 29.0\% & 23.7\% \\
	ConQAT Engine Core				 &  41 & 18.9\%  & 50.0\% & 40.5\% \\
	ConQAT Lib Commons				 &  45 &  9.5\%  & 56.3\% & 51.1\% \\
\midrule
	ConQAT dotnet				 			 & 154 & 36.3\%  & 48.1\% & 30.7\% \\
	Daisydiff						 			 &   7 &  6.4\%  & 49.8\% & 46.7\% \\
	Histone		  				 			 &  47 & 24.8\%  & 73.0\% & 54.9\% \\
	LittleProxy 				 			 &  35 & 71.4\%  & 45.4\% & 13.0\% \\
	Predictor						 			 &  80 & 52.7\%  & 72.4\% & 34.2\% \\
	Struts 2			 						 & 154 & 45.9\%  & 27.0\% & 14.6\% \\
	Symja								 			 & 234 & 25.0\%  & 21.3\% & 15.9\% \\
	Tspmccabe						 			 &  13 & 21.3\%  & 39.1\% & 30.7\% \\
\bottomrule
\end{tabular}
\label{Table_Results_Pseudo_Tested}
\end{table}

\begin{figure}
	\centering
	\includegraphics[width=\linewidth]{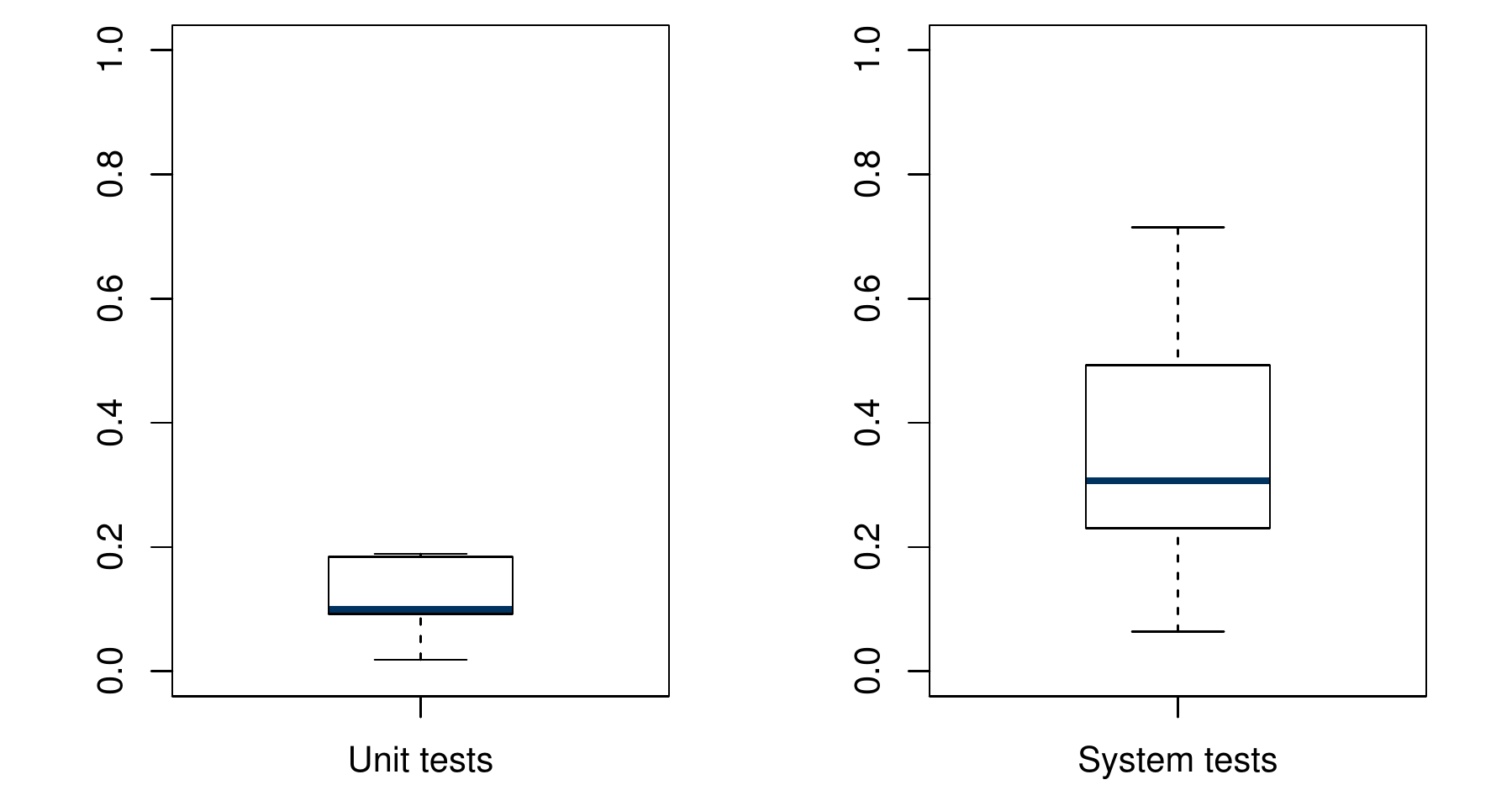}
	\caption{Boxplot comparing the ratios of pseudo-tested methods between unit and system tests}
	\label{Figure_RQ_1_2_Boxplot}
\end{figure}

\vspace{\baselineskip}
\noindent
\textbf{RQ 2: Does the ratio of pseudo-tested methods depend on the type of test?}
\\
\indent
The mean ratio of pseudo-tested methods of the study objects differs for unit and system tests.
The mean ratio was 11.41\% for unit tests and 35.48\% for system tests.
It is striking that the ratio is within a small range for unit tests
 but greatly deviates among systems tests.
The standard deviation of the ratio of pseudo-tested methods (6.42\% for unit and 20.60\% for system tests)
 confirms this observation.
The boxplot in Figure \ref{Figure_RQ_1_2_Boxplot} depicts the key figures and the deviation between unit and system tests.
The results show that the type of test influences the ratio of pseudo-tested methods.

\begin{figure*}
	\centering
	\includegraphics[width=\linewidth]{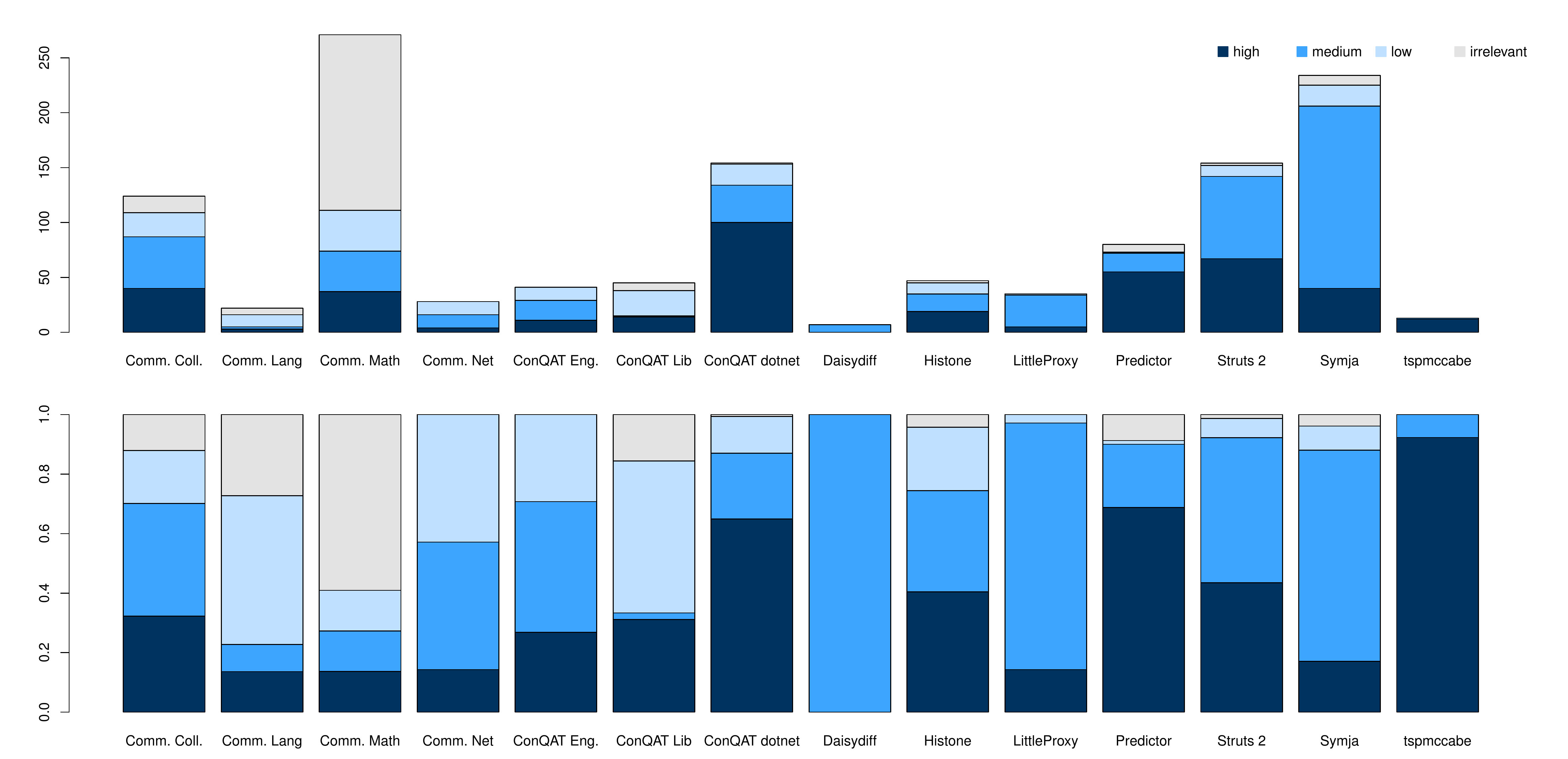}
	\caption{Pseudo-tested methods by their severity (top: absolute numbers, bottom: ratios)}
	\label{Figure_RQ_1_3_Barchart}
\end{figure*}

\vspace{\baselineskip}
\noindent
\textbf{RQ 3: How severe are the pseudo-tested methods?}
\\
\indent
Figure \ref{Figure_RQ_1_3_Barchart} presents the absolute and relative number of pseudo-tested methods, grouped by severity.
For 11 study objects, more than half of the pseudo-tested methods were of medium or high severity.
This was not the case for \textit{Apache Commons Math}, which contains a significant amount of irrelevant pseudo-tested methods
 (some test utility methods were mutated in the analysis); and \textit{Apache Commons Lang} as well as \textit{ConQAT Lib Commons} with some low-importance methods.
The results confirm the relevance of pseudo-tested methods and suggest that the lack of test effectiveness is a problem for a software project.

\section{Threats to Validity}
We separated the threats to validity into internal and external threats.

\subsection{Internal Threats}
The threats to internal validity comprise reasons why the results could be invalid for the study objects.

One threat to the internal validity is that some methods
 considered pseudo-tested might actually result in equivalent mutants.
We tried to mitigate this issue by the choice of the mutation operator and additional filtering.
The mutation operator modifies the whole method body,
 while many common operators only mutate single lines,
 and is therefore less likely to create an equivalent mutant.
Additionally, we filtered out empty and trivial methods such as one-line setters and getters (see Section \ref{Section_Mutation_Approach}).
As we generated two mutants for methods with primitive or string return types, the results were not distorted if only one mutants was equivalent.
We manually reviewed a random sample to make sure that the number of remaining undetected equivalent mutants was negligible.

Study objects with test cases that fail on the original code are a further threat to internal validity.
This can happen because of the test environment or faulty code in the study object.
Some test cases rely on further data stored in files, a database with a certain data model and content,
 other connected systems, or the network connection.
We tried to supply all the needed and available files in the test execution folder
 and set up the databases according to the project manuals.
Nevertheless, some test cases still failed.
This was the case for the \textit{Apache Commons Net} project, which presumably required certain firewall settings
 for some of its test cases. 
We excluded these failing test cases from the analysis.
If the excluded test cases had worked, they might have killed some mutants
 that were not killed by the other test cases (and therefore categorized as pseudo-tested).
For this reason, we only selected projects as study objects in which most test cases could be successfully executed on the original code.

Another threat to internal validity is custom class loaders that could interfere
 with the test execution on mutated code and affect the results in some seldom cases.
This may occur if a class is loaded multiple times by different class loaders during the execution of a single test case.
In this case, another class version is loaded in addition to the mutated one.
We consider this threat to be negligible.

One threat regarding the definition of test effectiveness is the fact that test cases can reveal faults that cause an exception to be thrown,
 even in pseudo-tested methods (that are considered as not effectively tested).

Concerning Research Question 3, the categorization of pseudo-tested methods according to their purpose was performed
 by considering the method name.
Due to the high number of methods, it was not feasible to inspect all the code
 to determine their purpose.
Therefore, some methods might actually belong to another category and severity than the assigned one.

\subsection{External Threats}
The external validity concerns the generalization of the results of the experiment.
One threat to external validity is that analyses were performed only for void methods
 and methods returning a primitive or string value.
Although we additionally executed the analysis for three (unit-test) study objects with methods returning objects
 and observed comparable results, the obtained results might not be representative of methods returning objects.

Furthermore, the results of the selected open-source pro\-jects might not be representative of closed-source systems.
We tried to mitigate this issue by considering several projects with different characteristics
 and application domains as study objects.
Further studies are necessary to determine if the results also apply to closed-source systems.

\section{Conclusions and Future Work}
The results of the experiment show that approximately 9\% to 19\% of the test-executed methods are pseudo-tested in projects with unit tests.
The ratio does not heavily deviate among study objects with unit tests (mean: 11.41\%, standard deviation: 6.42\%).
Therefore, code coverage at the method level is not completely misleading and can be used as an approximation for the effectiveness of unit tests.

In contrast, the ratio of pseudo-tested methods for system tests is generally higher than the ratio for unit tests.
It ranges between 6\% and 72\% for the analyzed study objects and heavily deviates (mean: 35.48\%, standard deviation: 20.60\%).
Therefore, code coverage at the method level is not a valid indicator for the effectiveness of system tests.

The assessment of the functional purpose and severity of pseudo-tested methods confirms the relevance of these methods for the software.
Faults in these methods would not be detected and could cause failures.

For future work, we intend to investigate characteristics of pseudo-tested methods and their relationship to test cases.
We want to find indicators that reveal pseudo-tested methods.
Such indicators would enable rapid detection of these methods in a static code analysis 
 and make the computationally expensive mutation testing approach superfluous.
This static code analysis could support developers in a continuously integrated development process.
Moreover, we plan to replicate the experiment with closed-source study objects.

\balance

\bibliographystyle{abbrv}

\begin{thebibliography}{10}

\bibitem{andrews2005mutation}
J.~H. Andrews, L.~C. Briand, and Y.~Labiche.
\newblock Is mutation an appropriate tool for testing experiments?
\newblock In {\em {Proc. 27th International Conference on Software Engineering
  (ICSE)}}. IEEE, 2005.

\bibitem{andrews2006using}
J.~H. Andrews, L.~C. Briand, Y.~Labiche, and A.~S. Namin.
\newblock Using mutation analysis for assessing and comparing testing coverage
  criteria.
\newblock {\em IEEE Transactions on Software Engineering}, 32(8), 2006.

\bibitem{daran1996software}
M.~Daran and P.~Th{\'e}venod-Fosse.
\newblock Software error analysis: a real case study involving real faults and
  mutations.
\newblock In {\em ACM SIGSOFT Software Engineering Notes}, volume~21. ACM,
  1996.

\bibitem{wwwFowler}
M.~Fowler.
\newblock {AssertionFreeTesting}.
\newblock \url{http://martinfowler.com/bliki/AssertionFreeTesting.html}.
\newblock Visited on December 21st, 2015.

\bibitem{grun2009impact}
B.~J. Gr{\"u}n, D.~Schuler, and A.~Zeller.
\newblock The impact of equivalent mutants.
\newblock In {\em Proc. International Conference on Software Testing,
  Verification and Validation Workshops (ICSTW)}. IEEE, 2009.

\bibitem{hutchins1994experiments}
M.~Hutchins, H.~Foster, T.~Goradia, and T.~Ostrand.
\newblock Experiments of the effectiveness of dataflow-and controlflow-based
  test adequacy criteria.
\newblock In {\em {Proc. 16th International Conference on Software Engineering
  (ICSE)}}. IEEE, 1994.

\bibitem{inozemtseva2014coverage}
L.~Inozemtseva and R.~Holmes.
\newblock Coverage is not strongly correlated with test suite effectiveness.
\newblock In {\em {Proc. 36th International Conference on Software Engineering
  (ICSE)}}. ACM, 2014.

\bibitem{marick1999misuse}
B.~Marick.
\newblock How to misuse code coverage.
\newblock \url{http://www.exampler.com/testing-com/writings/coverage.pdf}.
\newblock Visited on January 14th, 2016.

\bibitem{mockus2009test}
A.~Mockus, N.~Nagappan, and T.~T. Dinh-Trong.
\newblock Test coverage and post-verification defects: A multiple case study.
\newblock In {\em {Proc. 3rd International Symposium on Empirical Software
  Engineering and Measurement (ESEM)}}. IEEE, 2009.

\bibitem{namin2009influence}
A.~S. Namin and J.~H. Andrews.
\newblock The influence of size and coverage on test suite effectiveness.
\newblock In {\em Proc. 18th International Symposium on Software Testing and
  Analysis}. ACM, 2009.

\bibitem{niedermayr2013thesis}
R.~Niedermayr.
\newblock {Meaningful and practical measures for regression test reliability}.
\newblock Master's thesis, Technische Universit\"at M\"unchen, Munich, Germany,
  2013.

\bibitem{offutt1996experimental}
A.~J. Offutt, J.~Pan, K.~Tewary, and T.~Zhang.
\newblock An experimental evaluation of data flow and mutation testing.
\newblock {\em Software: Practice and Experience}, 26(2), 1996.

\bibitem{wong1994effect}
W.~E. Wong, J.~R. Horgan, S.~London, and A.~P. Mathur.
\newblock Effect of test set size and block coverage on the fault detection
  effectiveness.
\newblock In {\em Proc. 5th International Symposium on Software Reliability
  Engineering (ISSRE)}. IEEE, 1994.

\end{thebibliography}

\end{document}